\documentclass[3p,times]{elsarticle}
\usepackage{ecrc}
\volume{00}
\firstpage{1}
\journalname{Nuclear Physics A}
\runauth{A. Ficnar, J. Noronha, M. Gyulassy}
\jid{nupha}
\usepackage{amssymb}
\usepackage[figuresright]{rotating}

\begin{document}
\begin{frontmatter}

\title{Falling Strings and Light Quark Jet Quenching at LHC}

\author[a1]{Andrej Ficnar}
\author[a2]{Jorge Noronha}
\author[a1]{Miklos Gyulassy}
\address[a1]{Department of Physics, Columbia University, New York, NY 10027, USA}
\address[a2]{Instituto de F\'{i}sica, Universidade de S\~{a}o Paulo, 05315-970 S\~{a}o Paulo, Brazil}

\begin{abstract}
We explore phenomenological signatures of light quark jet quenching within the AdS/CFT correspondence. Numerical studies of the instantaneous energy loss of 
light quarks, modeled as falling strings, suggest a linear path dependence. We propose a phenomenological model for the energy loss and use it to compute the 
nuclear modification factor $R_{AA}$ for light quarks in an expanding plasma with Glauber initial conditions. The results are compared to the light hadron $R_{AA}$ 
data at the LHC and, although they show qualitative agreement, the quantitative disagreement we found motivated the exploration of effects from higher order derivative corrections to $AdS_5$ on this observable. 
\end{abstract}

\begin{keyword}
light quark energy loss, quark-gluon plasma, AdS/CFT correspondence
\end{keyword}

\end{frontmatter}

\section{Introduction}

Gauge/gravity duality has been an insightful tool in the study of many properties of the strongly-coupled quark-gluon plasma \cite{miklos-larry} created in heavy ion collisions at RHIC. The most studied example is the AdS/CFT correspondence \cite{maldacena,witten,gkb}: a duality between the $\mathcal{N}=4$ $SU(N_c)$ super-Yang-Mills theory and type IIB string theory on $AdS_5\times S^5$, which allows one to study this strongly coupled gauge theory in the $N_c\gg \lambda \gg 1$ limit by performing classical, two-derivative (super)gravity calculations. 

One of the important applications of the AdS/CFT correspondence has been the study of the phenomenon of jet quenching in strongly-coupled systems \cite{energylossrefs}. Experimental data on the suppresion of light hadrons in AA collisions from RHIC and LHC \cite{lhc1,lhc2} calls for a more consistent grasp on the energy loss of light quarks in the gauge/gravity duality in order to be able to compute jet quenching observables such as the nuclear modification factor $R_{AA}$ and the elliptic flow parameter $v_2$. In this work we will explore a possible way to compute the $R_{AA}$ using the input from the studies of the energy loss of light quarks in AdS/CFT and analyze its features.

\section{Light quarks in AdS/CFT and the instantaneous energy loss}

According to the AdS/CFT correspondence, a finite temperature $T$ in the plasma is dual to a black brane in the $AdS_5$ geometry with an event horizon at some radial coordinate $r_h \propto 1/T$ \cite{witten}. Degrees of freedom of mass $m_Q$ ('quarks') in the fundamental representation in the boundary theory correspond to a D7-brane in the $AdS$-BH geometry, which spans from $r=0$ (boundary) to some $r_m\sim 1/m_Q$ \cite{karch-katz}. Dressed quarks will then be dual to strings in the bulk with one or both endpoints on the D7-brane and the physics of the energy loss of these quarks will be directly related to the dynamics of their dual strings, which is, in the $N_c\gg \lambda \gg 1$ limit, classical. For light quarks, the D7-brane fills the entire $AdS$-BH geometry and a way to study their energy loss is to investigate the free motion of the strings that have both of their endpoints on the D7-brane (representing dressed $q\bar{q}$ pairs), the so-called falling strings \cite{chesler}.

In \cite{chesler} it was shown that the maximum stopping distance of light quarks in a strongly coupled $\mathcal{N}=4$ SYM plasma scales with energy as $\Delta x_{max}\sim E^{1/3}$. However, in general this information alone is not enough to calculate observables such as $R_{AA}$ or $v_2$, where the knowledge of the instantaneous energy loss is needed. To examine the instantaneous energy loss, one needs to analyze the spacetime momentum currents $\Pi_\mu^a$ on the string worldsheet, which, as demonstrated in \cite{chesler}, in case of falling strings become non-trivial, time-dependent quantities. 

In such a non-stationary environment, details of the geometry on the worldsheet become important, making the connection between the worldsheet currents and energy loss non-trivial. This was examined in \cite{my-paper}, where, by analyzing transformations of spacetime momentum fluxes on the classical string worldsheet, a general expression was derived for calculating the instantaneous energy loss in time-dependent string configurations valid in any choice of worldsheet parametrization. That expression shows that the energy loss in time-dependent string configurations receives a correction to the simple $\Pi_\mu^\sigma$ component of the flux. This correction comes from the fact that the points on the string at which we want to evaluate the energy loss at different times (and which serve to define what we mean by a 'jet') do not necessarily have constant coordinates in the chosen worldsheet parametrization.

In the case of falling strings and defining the jet as a part of the string within a certain fixed $\Delta x \sim 1/(\pi T)$ distance from the endpoint (as in \cite{chesler}), it was demonstrated \cite{my-paper} that this correction becomes especially important at late times and substantially decreases the magnitude of the Bragg-like peak reported in \cite{chesler}. In fact, as noted in \cite{my-paper}, preliminary numerical studies suggest that, although the early time behavior of the energy loss is susceptible to the initial conditions, the linearity of it, $dE/dt\sim t$, seems to be a remarkably robust feature, which we will use as a 
working assumption in the calculation of $R_{AA}$. We point out, however, that the time behavior of the instantaneous energy loss is related to the specific jet definition used and, in principle, changing that definition can affect the time dependence of the energy loss.

\section{Calculating $R_{AA}$}
Motivated by \cite{miklos-abc}, we model the phenomenologically relevant part of the instantaneous energy loss of a light quark moving through an $\mathcal{N}=4$ SYM plasma at a temperature $T$ in a following generic way:
\begin{equation}\label{eq1}
\frac{dE}{dx}(E_0,T,x)=-c(E_0,T)x^1\Theta\left[L_s(E_0,T)-x\right]\, .
\end{equation}
Here $x$ is the distance in the medium the quark has traversed, $E_0$ is its initial energy, $L_s$ is the stopping distance (which will be defined shortly) and $\Theta$ is the step function. The initial energy and temperature dependence has been packed into an unknown function $c(E_0,T)$ and the linear path dependence, as noted in the previous section, was inferred from the preliminary numerical studies of falling strings in $AdS_5$. The second piece of information we need is the dependence of the stopping distance on the initial energy and temperature which we know from \cite{chesler}:
\begin{equation}\label{eq2}
L_s(E_0,T)=\frac{\kappa}{T} \left(\frac{E_0}{\sqrt{\lambda}T}\right)^{1/3}\, .
\end{equation}
Here $\lambda$ is the 't Hooft coupling and $\kappa$ is a numerical factor that depends on the initial conditions of the string (for the maximum stopping distance it was shown to be $\approx 0.5$ \cite{chesler}). Requiring that the energy lost over the stopping distance must be the quark's initial energy fixes the unknown function $c(E_0,T)$ and we arrive at:
\begin{equation}\label{eq5}
\frac{dE}{dx}=-\chi E_0^{1/3}x^1T^{8/3}\, ,
\end{equation}
where we have defined an effective coupling $\chi\equiv 2\lambda^{1/3}/\kappa^2$, which determines the overall magnitude of the energy loss. As noted in \cite{my-paper}, formula (\ref{eq5}) has a surprising similarity to the typical qualitative behavior of energy loss of light quarks in pQCD in the strong LPM regime \cite{miklos-abc}. 

Using this formula, we can proceed to compute the $R_{AA}$ for light quarks, in the same way it was done in \cite{proceeding2}, using the Glauber initial conditions which determine the temperature profile of the expanding plasma and averaging over the jet azimuthal directions and production points in the transverse plane. The results are shown in the left plot in Fig.\ 1. Note that the only free parameter in (\ref{eq5}) is the effective coupling $\chi=2\lambda^{1/3}/\kappa^2$, which we vary in Fig.\ 1. From \cite{chesler} we know that $\kappa\approx 0.5$ so, with even an unphysically small $\lambda=1$, we need a minimal value of $\chi\approx 8$. From the left plot in Fig. 1, we see that such a value of $\chi$ gives an $R_{AA}$ of a rather low magnitude, indicating strong quenching. If one uses even lower values of $\chi$, we see in Fig. 1 that indeed $R_{AA}$ has the correct qualitative behavior as displayed by the LHC data on the suppression of light hadrons in AA collisions \cite{lhc1,lhc2}. This suggests that the main problem here could be simply in the low magnitude of $R_{AA}$, or, equivalently, that the quenching is too strong.

\begin{figure}[h!]
	  \centering
    \includegraphics[width=51mm]{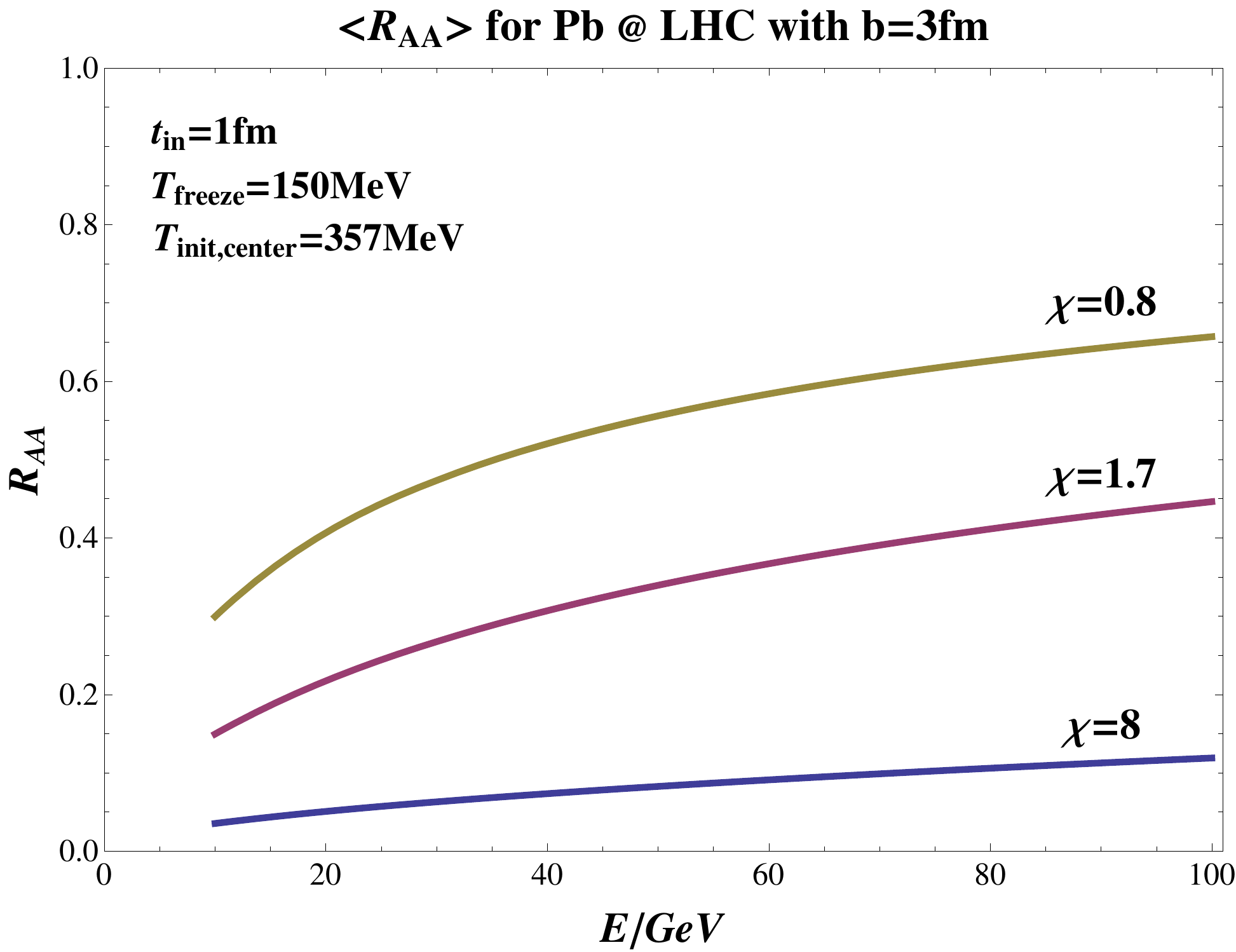}
    \includegraphics[width=51mm]{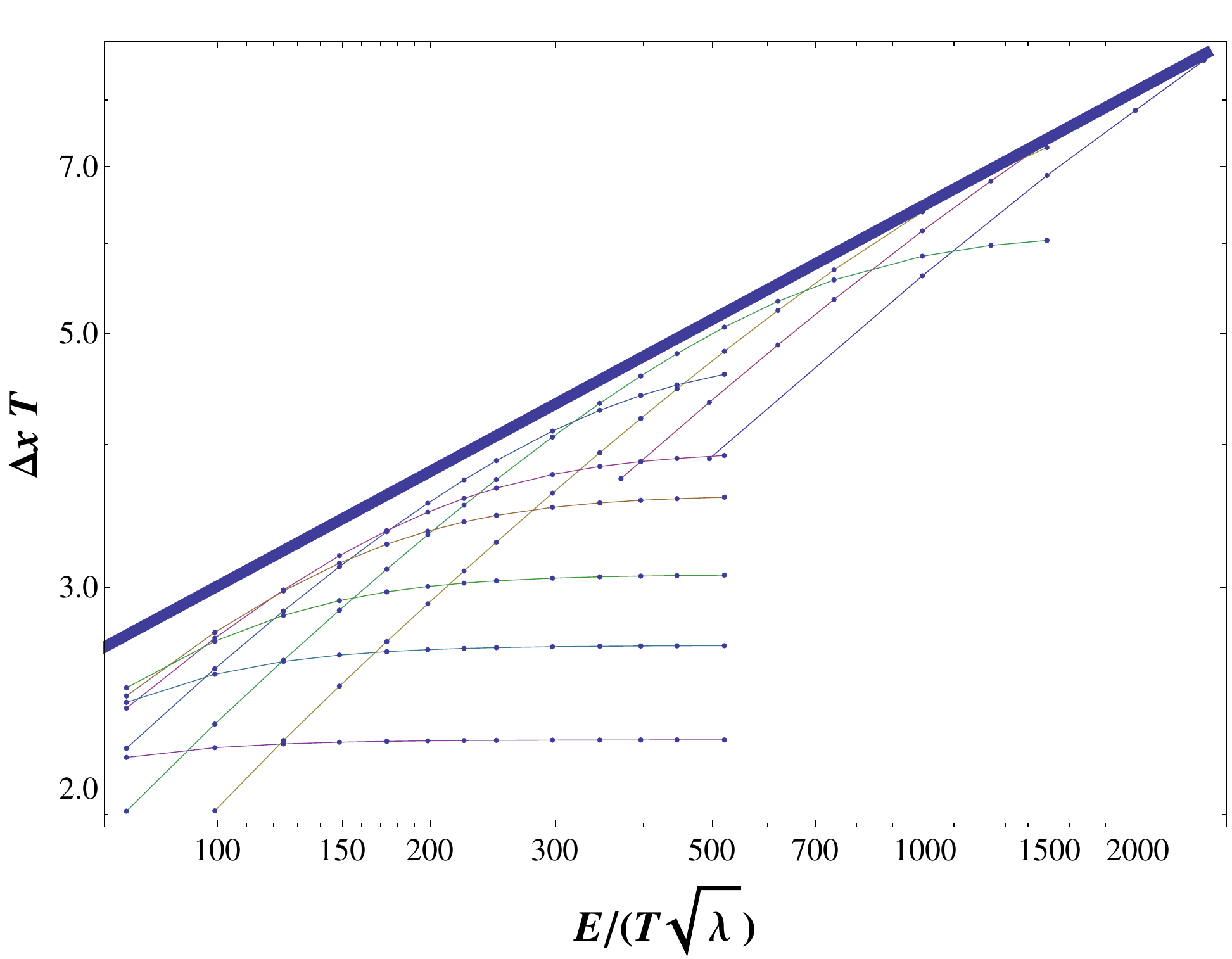}
    \includegraphics[width=51mm]{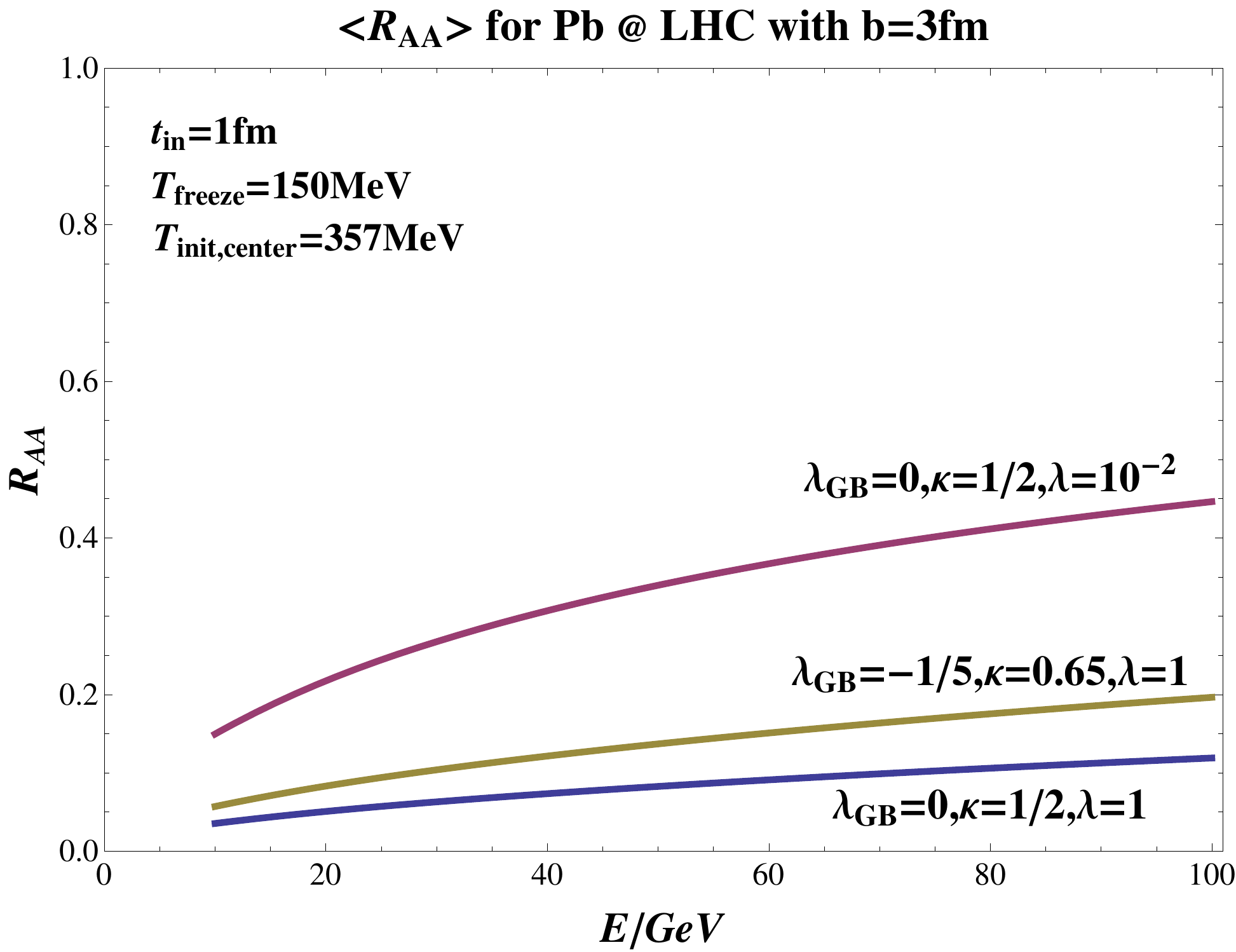}\\
    \vskip 0.1in
    \parbox[l]{162mm}{\footnotesize{\textbf{Figure 1.}} \textit{Left:} Nuclear modification factor $R_{AA}$ for light quarks in 
central Pb+Pb collisions at LHC as a function of the final parton energy $E$ for several different values of the effective coupling $\chi$. 
The initial jet production time $t_i$, the freezout temperature $T_{freeze}$ and the initial temperature at the center of the plasma $T_{init,center}$ 
are indicated in the plot. \textit{Center:} Stopping distance $\Delta x$ of falling strings as a function of the 
initial string energy $E$ for different initial conditions (dots connected by lines represent the same family of initial conditions) 
in the $AdS_5-GB$ geometry (\ref{eq7}) for $\lambda_{GB}=-1/5$. The thick blue line is the numerically extrapolated 
envelope $\Delta x T=0.65(E/(\sqrt{\lambda}T))^{1/3}$. \textit{Right:} Comparison of the light quark $R_{AA}$ at LHC as a function of the final parton 
energy $E$, computed as in \cite{proceeding2}, for different values of the effective coupling $\chi$ with and without Gauss-Bonnet higher 
derivative corrections (\ref{eq7}).}
\end{figure}

\section{Higher derivative corrections}
We decided to examine if higher derivative corrections to the gravity sector of $AdS_5$ can decrease the strength of the quenching while keeping the correct qualitative behavior of $R_{AA}$. In the presence of a $D7$-brane, the leading $1/N_c$ corrections come from the $R^2$-terms which we will model by a Gauss-Bonnet term, i.e. we will consider the Lagrangian density of the form $\mathcal{L}=R+\frac{12}{L^2}+L^2\frac{\lambda_{GB}}{2}\left(R^2-4R_{\mu\nu}^2+R_{\mu\nu\rho\sigma}^2\right)$. There, $\lambda_{GB}$ is a dimensionless parameter, which is constrained to be $-\frac{7}{36}<\lambda_{GB}\leq \frac{9}{100}$ to avoid causality violations \cite{myers} and to have positive energy density on the boundary \cite{hofman-maldacena}. A black brane solution in this case is known analytically \cite{cai}:
\begin{equation}\label{eq7}
ds^2=\frac{L^2}{r^2}\left[-a^2f_{GB}(r)dt^2+d\vec{x}^2+\frac{dr^2}{f_{GB}(r)}\right]\, ,
\end{equation}
where $f_{GB}(r)=\frac{1}{2\lambda_{GB}}\left[1-\sqrt{1-4\lambda_{GB}\left(1-r^4/r_h^4\right)}\right]$ and $a^2=\frac{1}{2}\left(1+\sqrt{1-4\lambda_{GB}}\right)$ is here to make the speed of light at the boundary ($r=0$) equal to one. In this geometry the relations of the 't Hooft coupling and temperature in the boundary theory to the fundamental string length and parameter $r_h$, respectively, are altered: $\sqrt{\lambda}=\frac{L^2 a^2}{\alpha'}$ and $T=\frac{a}{\pi r_h}$. 

Following a similar procedure as in \cite{chesler}, by analyzing null geodesics in geometry (\ref{eq7}) and relating its parameters to the energy of the string, we have the following preliminary result for the maximum stopping distance of falling strings up to linear order in $\lambda_{GB}$ \cite{our-future-paper}:
\begin{equation}\label{eq8}
\Delta x_{max}=\frac{\mathcal{C}}{T}\left(\frac{E}{T\sqrt{\lambda}}\right)^{1/3}\left(1-\mathcal{F}\lambda_{GB}\right)+\mathcal{O}(\lambda_{GB}^2)\, ,
\end{equation}
where $\mathcal{C}\sim 1/2$ and $\mathcal{F}$ is a numerical factor $\gtrsim 1$, for which our preliminary estimate gives $\mathcal{F}=11/6$. We see that the $\sim E^{1/3}$ scaling is still present and that for negative values of $\lambda_{GB}$ we can increase the stopping distance by up to $\sim 30-40$\%, compared to the case of pure $AdS_5$ with no higher derivative corrections. The origin of the non-negligible factor $\mathcal{F}$ is, in a sense, a collection of small effects: the null geodesic stopping distance in geometry (\ref{eq7}) is larger, the temperature and the 't Hooft coupling both depend on $\lambda_{GB}$, and these expressions, when expanded in $\lambda_{GB}$, all add up in the same direction to give this factor. In addition to the $\sim E^{1/3}$ scaling, there are indications that, at linear order in $\lambda_{GB}$, we can also have an additional new energy scaling (though numerically suppressed), details of which, together with the precise value of the parameter $\mathcal{F}$ in (\ref{eq7}), will be presented in a future publication \cite{our-future-paper}.

Since the Gauss-Bonnet parameter $\lambda_{GB}$ affects the $\kappa$ coefficient in (\ref{eq2}) in non-trivial manner, we can expect to see the effects of this in the $R_{AA}$: the numerical results are shown in Fig.\ 1, center and right plots. From the center plot we see that the numerically extrapolated $\sim 0.65\, E^{1/3}$ envelope agrees well with the maximum stopping distance scaling from formula (\ref{eq8}) with $\mathcal{F}=11/6$, which predicts the $\kappa$ coefficient of $\approx 0.68$. The sensitivity of $R_{AA}$ to this is shown in the right plot of Fig.\ 1, where we see that keeping $\lambda=1$ fixed, just by ``turning on'' the higher derivative corrections, we can increase $R_{AA}$ by almost $100$\%.

\section{Conclusions, Outlook, and Acknowledgments}

Using a phenomenological model (\ref{eq5}) for the instantaneous energy loss of light quarks in AdS/CFT, we have calculated the nuclear modification factor $R_{AA}$ for light quarks in an expanding plasma with Glauber initial conditions \cite{proceeding2}. While our model calculations for $R_{AA}$ qualitatively agree with the LHC light hadron $R_{AA}$ data \cite{lhc1,lhc2}, the overall magnitude is too low. We have suggested that taking into account higher derivative corrections ($R^2$) to the gravity sector of $AdS_5$ should increase the stopping distance (\ref{eq8}) by a non-trivial factor to which $R_{AA}$ is quite sensitive (Fig.\ 1).

It would be interesting to study the effect of higher derivative corrections on the energy loss of light quarks embedded in a non-conformal plasma (the energy loss of heavy quarks in simple non-conformal models was studied in \cite{proceeding1,proceeding2}). It would also be interesting to examine the falling strings in the Janik-Peschanski metric \cite{jp}, which is dual to perfect fluid hydrodynamics, to take into account effects from the rapid expansion of the strongly-coupled plasma \cite{our-future-paper}.

We thank W. Horowitz, A. Buzzatti, G. Torrieri and S. S. Gubser for helpful discussions. A.F. and M.G. acknowledge support by US-DOE Nuclear Science Grant No. DE-FG02-93ER40764. J.N. is supported by the Brazilian funding agencies FAPESP and CNPq.  

\bibliographystyle{elsarticle-num}

\end{document}